\documentclass[12pt]{article}
\usepackage{epsfig}

\def\beq{\begin{equation}}
\def\eeq#1{\label{#1}\end{equation}}
\def\eeqn{\end{equation}}

\def\beqa{\begin{eqnarray}}
\def\eeqa#1{\label{#1}\end{eqnarray}}
\def\eeqan{\end{eqnarray}}

\let\bar=\overbar

\def\Dslash{\not{\hbox{\kern-4pt $D$}}}
\def\dslash{\not{\hbox{\kern-2pt $\del$}}}

\def\msb{{\bar{\ssstyle M \kern -1pt S}}}

\usepackage{wrapfig,rotating}
\usepackage{amssymb}
\usepackage{amsmath}
\newcommand{\PO}{\rm l \! P }

\newcommand{\xpom}{x_{\PO} }

\newcommand{\be}{\begin{equation}}

\def\Title#1{\begin{center} {\Large {\bf #1} } \end{center}}

\begin{document}

\Title{Diffraction at HERA}

\bigskip\bigskip

%+\addtocontents{toc}{{\it D. Reggiano}}
%+\label{ReggianoStart}

\begin{raggedright}  

{\it Laurent Schoeffel\index{Schoeffel, L.}\\
CE Saclay\\
Irfu/SPP\\
F-91191 Gif-sur-Yvette Cedex, FRANCE}
\bigskip\bigskip
\end{raggedright}

\section{Introduction}

Between 1992 and 2007, the 
HERA accelerator provided $ep$ collisions at center
of mass energies beyond $300 \ {\rm GeV}$
at the interaction points of the H1 and ZEUS experiments. 
Perhaps the most interesting results to emerge relate to
the newly accessed field of 
perturbative strong interaction physics at low Bjorken-$x$,
where parton densities become extremely large.
Questions arise as to how and where non-linear dynamics tame
the parton density growth \cite{saturation} 
and challenging features such as geometric scaling \cite{geo:scale}
are observed. Central to this low $x$ physics landscape 
is a high rate of diffractive
processes, in which a colorless exchange takes place and
the proton remains intact. 
In particular, the study of semi-inclusive diffractive
deep-inelastic scattering (DDIS),
$\gamma^* p \rightarrow X p$ \cite{hera:diff,Schoeffel:2010bb} 
has led to a revolution in our microscopic, parton level,
understanding of the structure of elastic
and quasi-elastic high energy hadronic scattering. 
Comparisons
with hard diffraction in proton-(anti)proton scattering have
also improved our knowledge of absorptive 
and underlying event effects in which the 
diffractive signature may be obscured by multiple interactions in the
same event \cite{gap:survival}. In addition to their fundamental
interest in their own right, these issues are highly relevant
to the modeling of chromodynamics at the LHC \cite{lhcdiff}.  
 
\begin{figure}[h]
\centerline{\hspace*{0.1cm}
            {\Large{\bf{(a)}}}
            \includegraphics[height=0.3\textwidth]{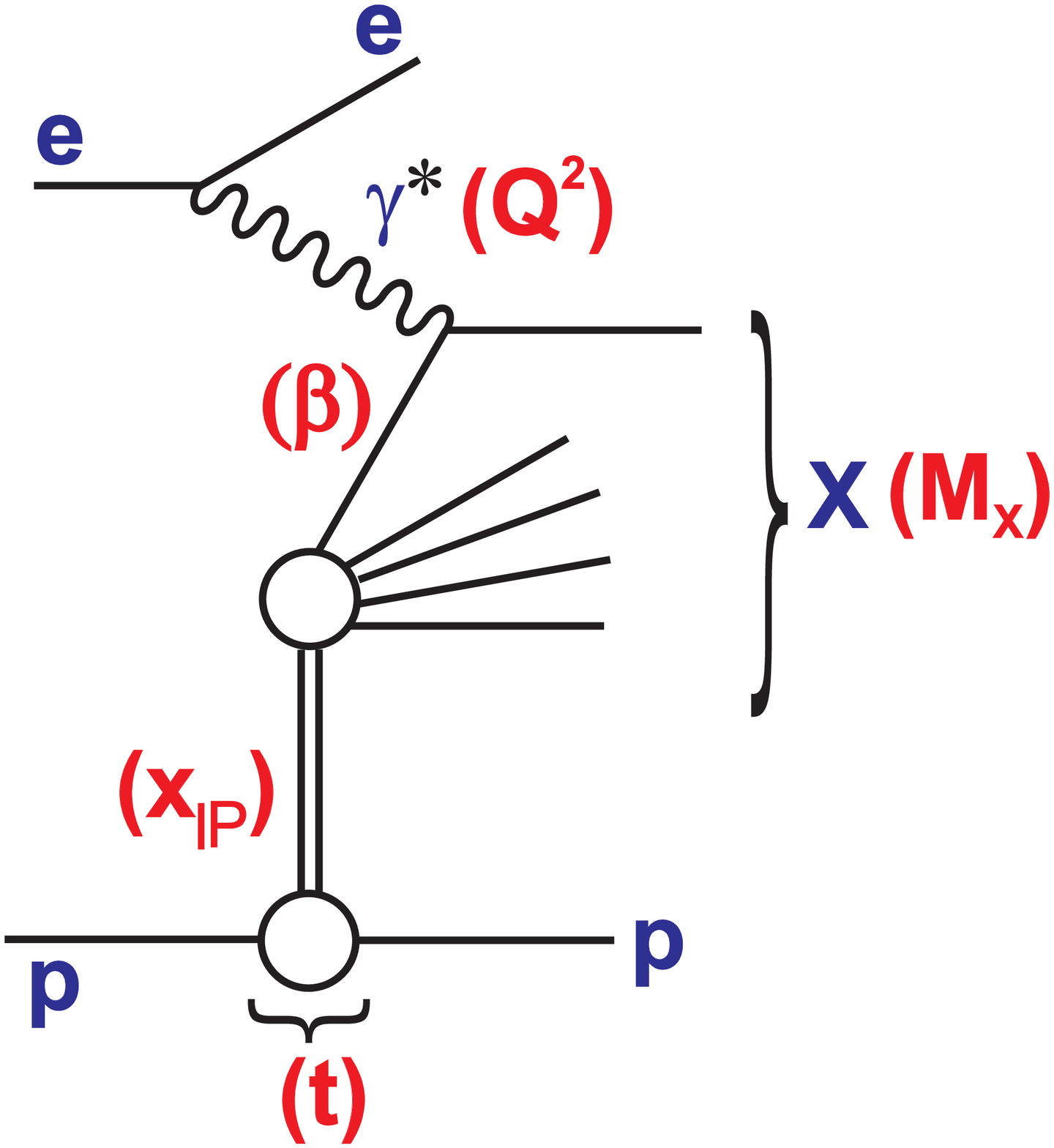}
            \hspace*{0.1cm}
            {\Large{\bf{(b)}}}
            \includegraphics[height=0.3\textwidth]{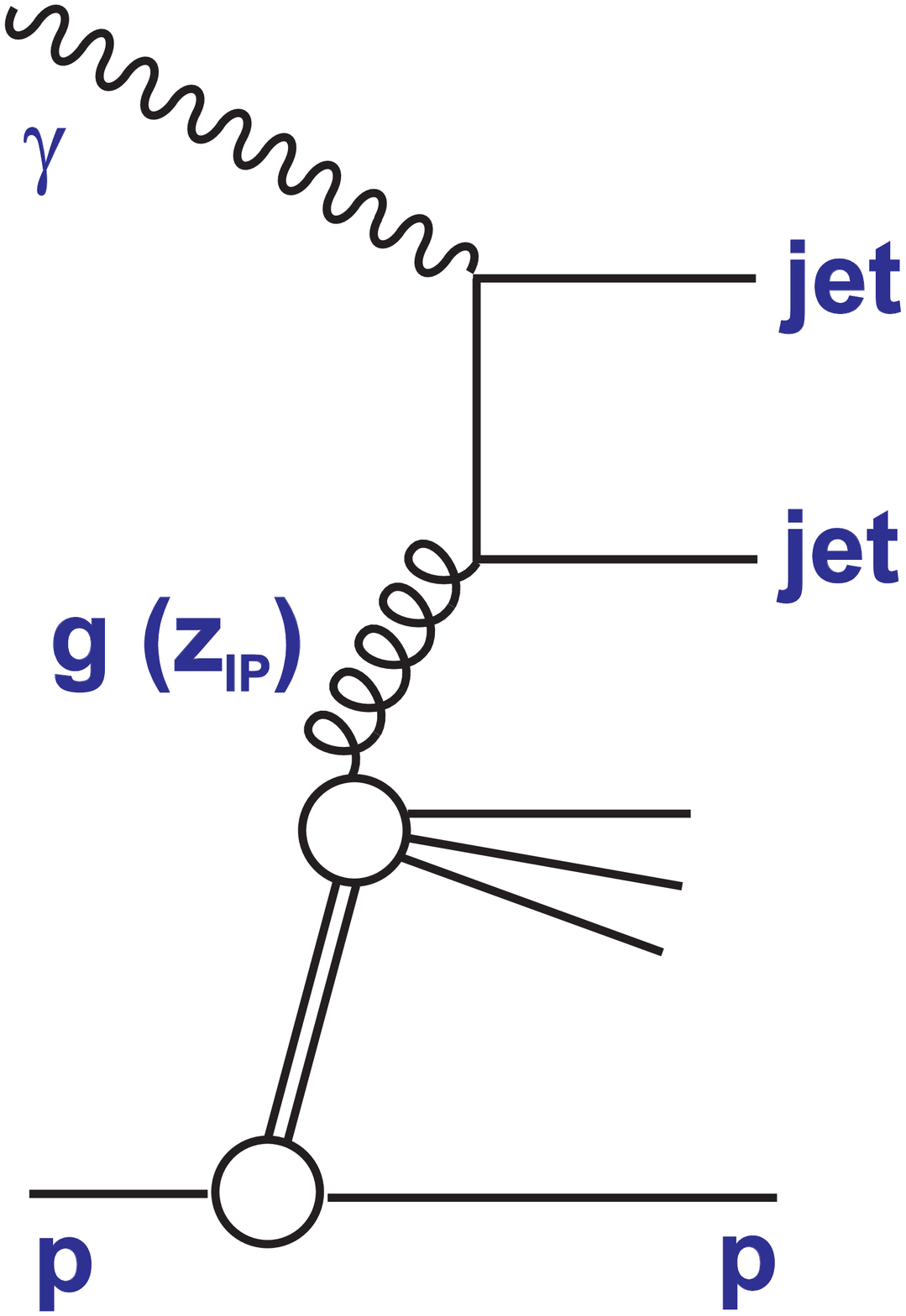}
            \hspace*{0.6cm}
            {\Large{\bf{(c)}}}
            \includegraphics[height=0.3\textwidth]{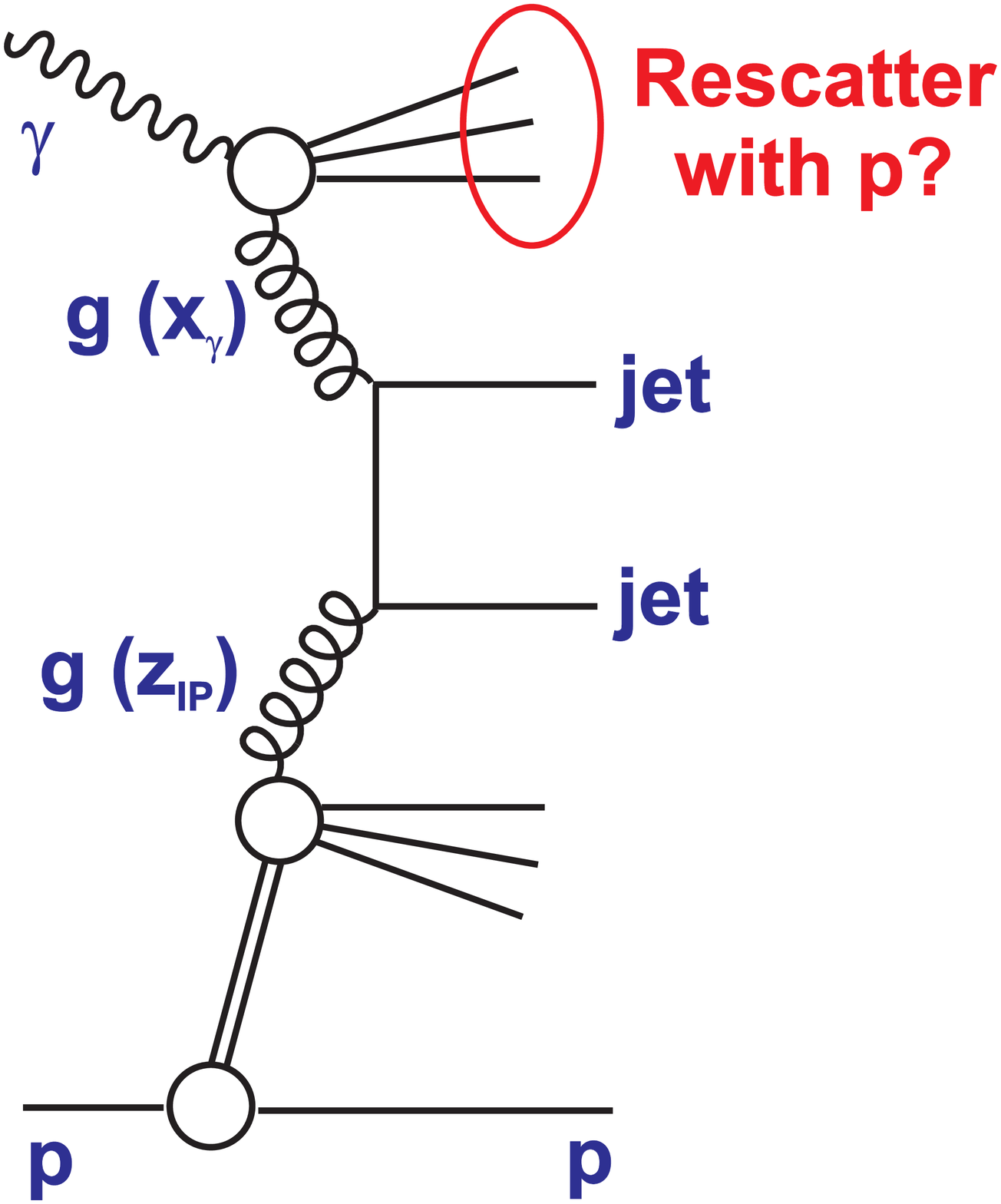}}
\caption{Sketches of diffractive $ep$ processes.
(a) Inclusive DDIS at the level of the quark parton model,
illustrating the kinematic variables 
discussed in the text. (b) Dominant leading order diagram for 
hard scattering in DDIS or direct photoproduction, in
which a parton of momentum fraction $z_{I\!\!P}$ from the DPDFs
enters the hard scattering. (c) A leading
order process in resolved photoproduction involving a parton 
of momentum fraction $x_\gamma$ relative to the photon.}
\label{feynman}
\end{figure}

The kinematic variables describing 
DDIS are illustrated in Fig.\ref{feynman}a.
The longitudinal momentum fractions
of the colorless exchange with respect to the incoming
proton and of the struck quark with respect to the colorless
exchange are denoted $x_{_{I\!\!P}}$ and $\beta$, respectively,
such that $\beta \, x_{_{I\!\!P}} = x$.
The squared four-momentum transferred at
the proton vertex is given by the Mandelstam $t$ variable.
The semi-inclusive DDIS cross section is usually presented in 
the form of a diffractive reduced cross section $\sigma_r^{D(3)}$,
integrated over $t$ and related to the experimentally measured
differential cross section by \cite{h1:lrg}
\begin{equation}
\frac{{\rm d}^3\sigma^{ep \rightarrow e X p}}{\mathrm{d} x_{_{I\!\!P}} \ \mathrm{d} x \ \mathrm{d} Q^2} = \frac{2\pi
  \alpha^2}{x Q^4} \cdot Y_+ \cdot \sigma_{r}^{D(3)}(x_{_{I\!\!P}},x,Q^2) \ ,
\label{sigmar}
\end{equation}
where $Y_+ = 1 + (1-y)^2$ and $y$ is the usual Bjorken variable.
%Similarly to inclusive DIS, 
The reduced cross section depends 
at moderate scales, $Q^2$, on two
diffractive structure functions
$F_2^{D(3)}$ and $F_L^{D(3)}$ according to
\begin{equation}
\sigma_r^{D(3)} =
F_2^{D(3)} - \frac{y^2}{Y_+} F_L^{D(3)}.
\label{sfdef}
\end{equation}
For $y$ not too close to unity,
$\sigma_r^{D(3)} = F_2^{D(3)}$ holds to very good approximation.

\section{Measurement methods and comparisons}

Experimentally, diffractive $ep$ scattering
is characterized by the presence of a leading proton in the
final state, retaining most of the initial state proton energy, and
by a lack of hadronic activity in the
forward (outgoing proton) direction, such that the
system $X$ is cleanly separated and 
its mass $M_X$ may be measured in the central
detector components.
These signatures have been widely exploited at HERA to select
diffractive events by tagging the outgoing proton
in the H1 Forward Proton Spectrometer or
the ZEUS Leading Proton Spectrometer 
(`LPS method' \cite{h1:fps,zeus:lrglps,h1:fpshera2}) or
by requiring the presence of a large gap in the rapidity
distribution of hadronic final state particles
in the forward region 
(`LRG method' \cite{h1:lrg,zeus:lrglps,H1:newdata}).
In a third approach, not considered in detail here, 
the inclusive DIS sample is
decomposed into diffractive and non-diffractive contributions based
on their characteristic dependences on $M_X$ \cite{H1:newdata,zeus:mx}.
Whilst the LRG and $M_X$-based techniques yield better
statistics than the LPS method, they suffer from systematic uncertainties 
associated with an admixture 
%at the $20 \%$ level 
of proton dissociation to low mass states, which 
is irreducible due to the limited forward detector acceptance.

\begin{figure}[h]
\centerline{\includegraphics[width=0.4\textwidth]{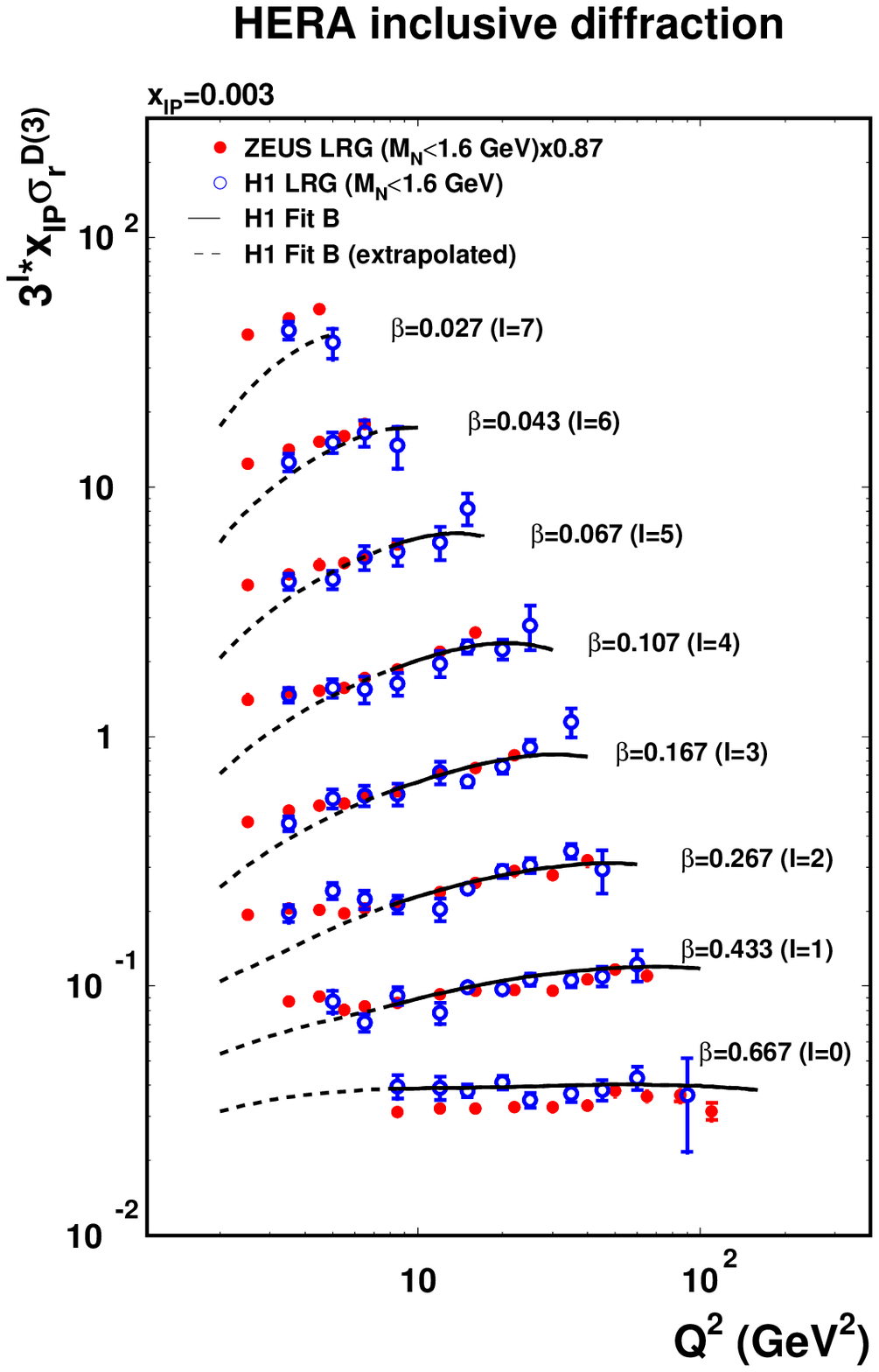}
            \hspace*{1cm}
            \includegraphics[width=0.4\textwidth]{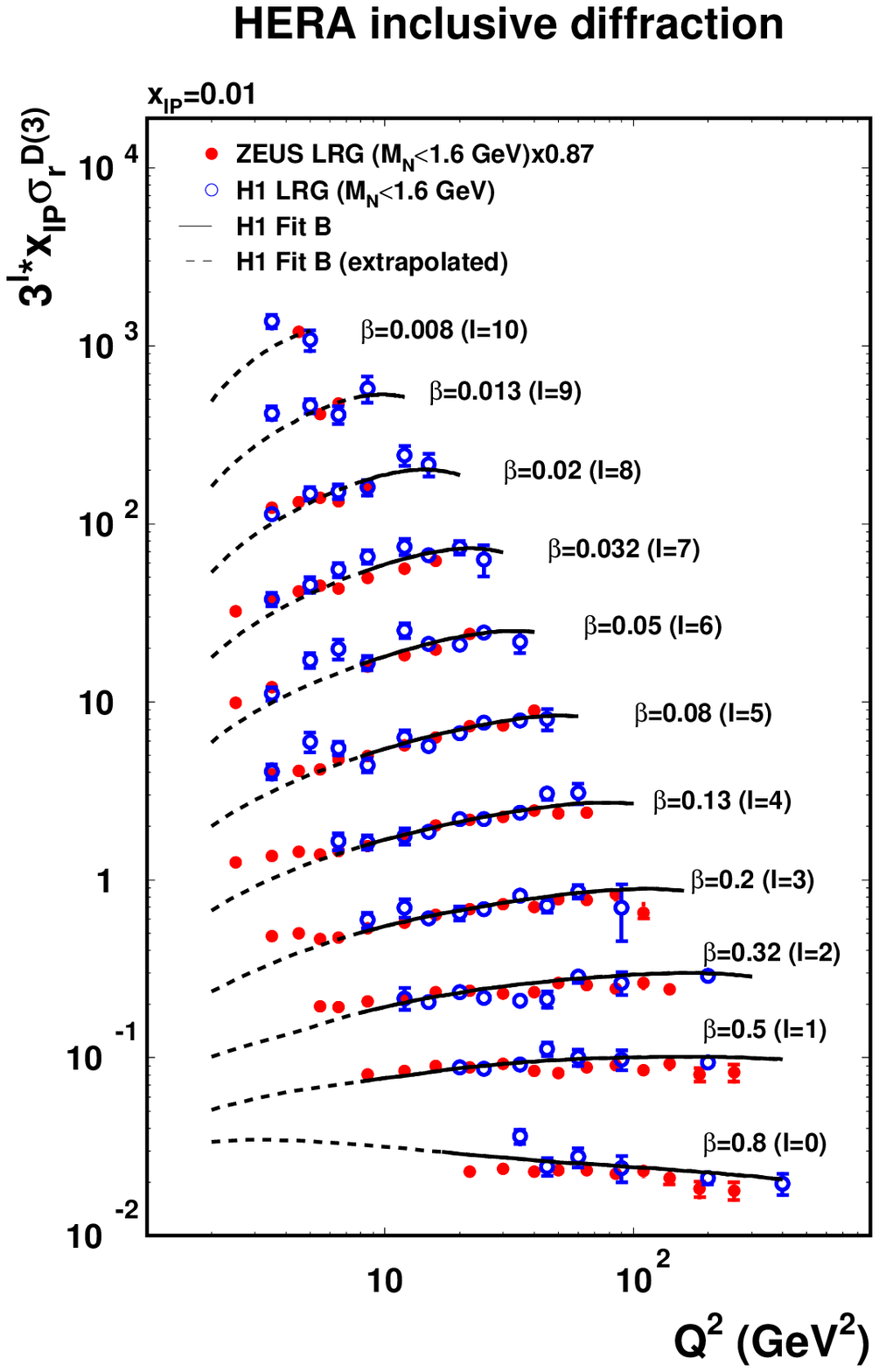}}
\caption{H1 and ZEUS measurements of the diffractive reduced cross section
%The $Q^2$ dependence is shown at numerous $\beta$ values 
at two example $x_{I\!\!P}$ values \cite{newman:ruspa}. 
The ZEUS data are
scaled by a factor of $0.87$ to match the H1 normalisation. 
The data are compared with the results of the  
H1 2006 Fit B DPDF based parameterization \cite{h1:lrg}
for $Q^2 \geq 8.5 \ {\rm GeV^2}$
and with its DGLAP (QCD) based extrapolation to lower $Q^2$.}
\label{h1vzeus}

\end{figure}

\begin{figure}[htbp]
\centerline{\includegraphics[width=1.\textwidth]{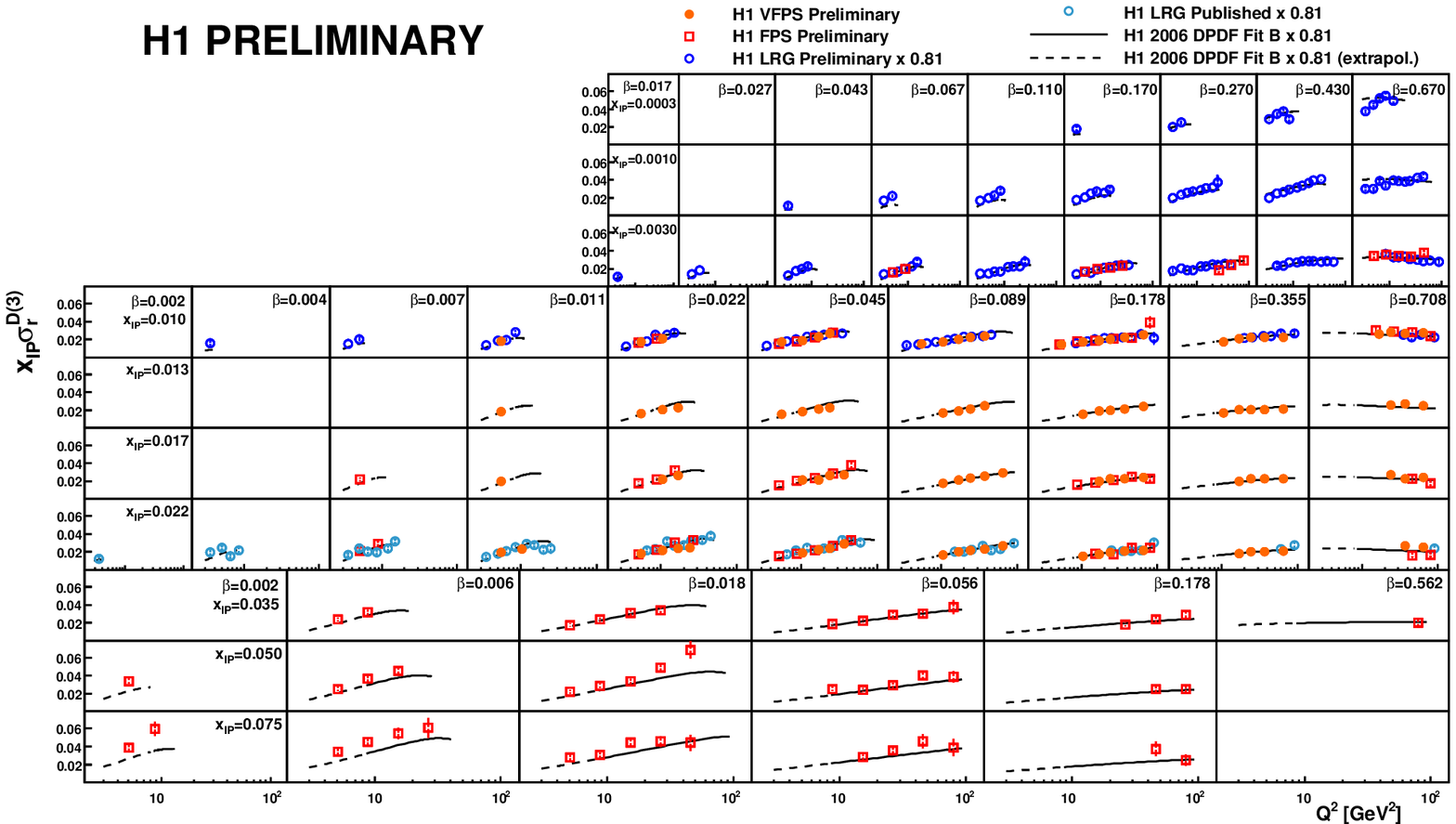}}
\caption{H1 measurements of the diffractive reduced cross section.
The $Q^2$ dependence is shown at numerous $\beta$ and $\xpom$ values.}
\label{h1vzeusb}
\end{figure}
\begin{figure}[htbp]
\centerline{\includegraphics[width=1.\textwidth]{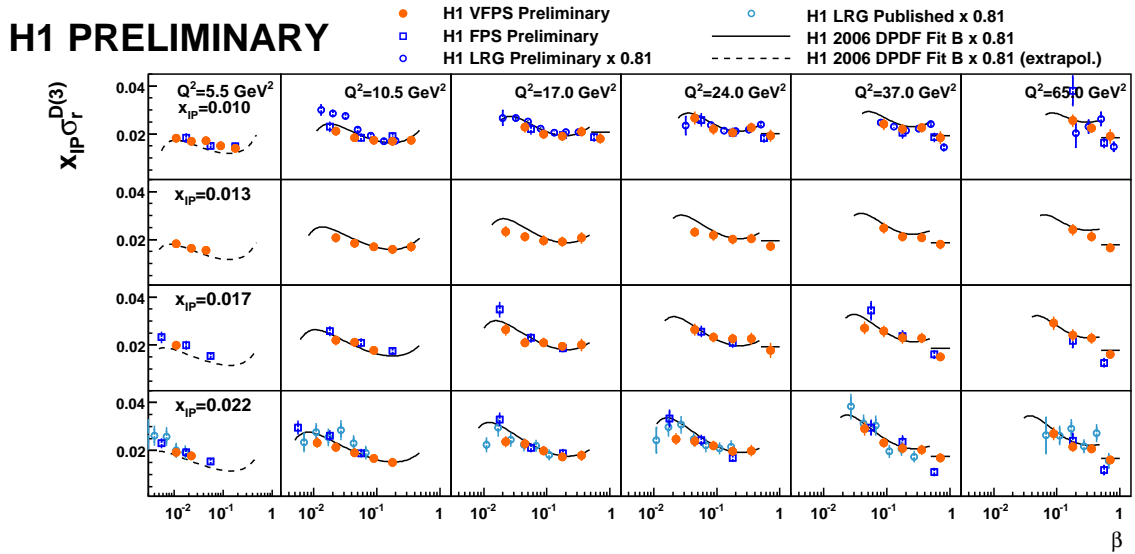}}
\caption{H1 measurements of the diffractive reduced cross section.
The $\beta$ dependence is shown at numerous $Q^2$ and $\xpom$ values.}
\label{h1vzeusc}
\end{figure}

The H1 collaboration recently released a preliminary 
proton-tagged measurement
using its full available FPS sample at HERA-II \cite{h1:fpshera2}. 
The integrated luminosity
is $156 \ {\rm pb^{-1}}$, a factor of 20 beyond previous H1
measurements. The new data tend to lie slightly above
the recently published 
final ZEUS LPS data from HERA-I \cite{zeus:lrglps}, but are
within the combined normalization uncertainty of around $10\%$.
The most precise test of compatibility between H1 and ZEUS is
obtained from the LRG data. The recently published ZEUS data \cite{zeus:lrglps}
are based on an integrated luminosity of $62 \ {\rm pb^{-1}}$ and
thus have substantially improved statistical precision compared
with the older H1 published results \cite{h1:lrg}. The normalization 
differences between the two experiments are most obvious here,
having been quantified at $13\%$, which is a little beyond
one standard deviation in the combined normalization uncertainty.
After correcting for this factor, very good agreement is observed
between the shapes of the H1 and ZEUS cross sections throughout most of
the phase space studied, as shown in 
Fig. \ref{h1vzeus}.
A more detailed comparison between different diffractive cross section
measurements by H1 and ZEUS and a first attempt to combine the 
results of the two experiments can be found in \cite{newman:ruspa}.
Fig. \ref{h1vzeusb} and \ref{h1vzeusc} present a complete summary of various measurements of the H1
experiment (using different experimental methods).

\section{Soft physics at the proton vertex}
\label{soft}

\begin{figure}[h]
\centerline{\hspace*{0.3cm}
            {\Large{\bf{(a)}}}
            \includegraphics[width=0.45\textwidth]{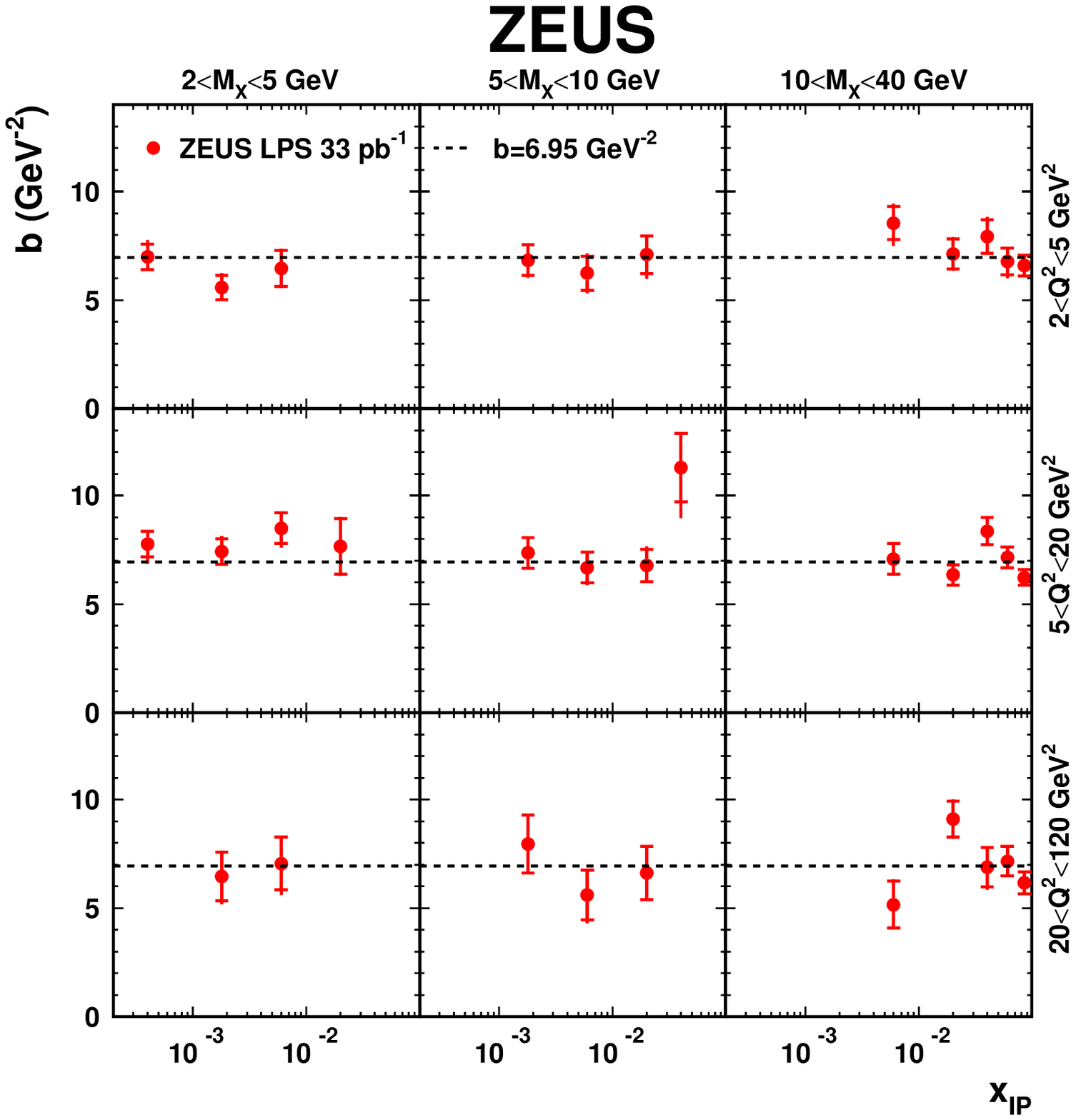}
            \hspace*{0.3cm}
            {\Large{\bf{(b)}}}
            \hspace*{-0.3cm}
            \includegraphics[width=0.45\textwidth]{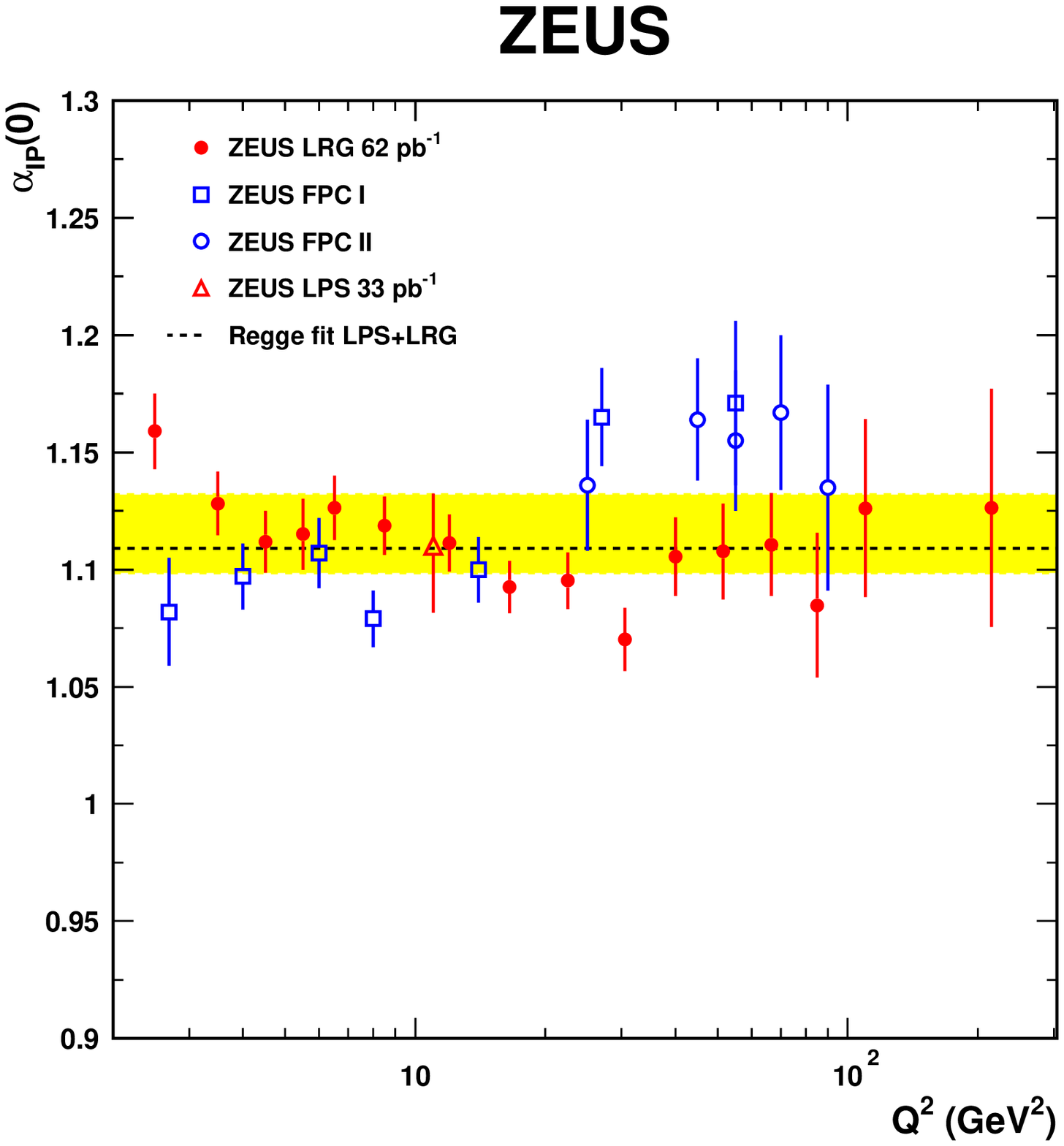}}
\caption{a) Measurements of the exponential $t$ slope from ZEUS
LPS data, shown as a function of $Q^2$, $x_{I\!\!P}$ and $M_X$.
b) ZEUS extractions of the effective pomeron intercept describing
the $x_{I\!\!P}$ dependence of DDIS data at different $Q^2$ 
values \cite{zeus:lrglps}.}
\label{regge}
\end{figure}

To good approximation, LRG and LPS 
data show \cite{h1:lrg,h1:fps,zeus:lrglps} that 
DDIS data satisfy a proton vertex 
factorization,
whereby 
the dependences on variables which describe the scattered 
proton ($x_{I\!\!P}$, $t$) factorise from those
describing the hard partonic interaction ($Q^2$, $\beta$).
For example, 
the slope parameter $b$, extracted in \cite{zeus:lrglps}
by fitting the $t$ distribution to the form
${\rm d} \sigma / {\rm d} t \propto e^{b t}$, is shown as a function
of DDIS kinematic variables in Fig.~\ref{regge}a. 
There are no significant variations
from the average value of $b \simeq 7 \ {\rm GeV^{-2}}$ anywhere in the
studied range.   
The measured value of $b$ is significantly larger than
that from `hard' 
exclusive vector meson production ($ep \rightarrow eVp$). 
It is 
characteristic of an interaction region 
of spatial extent considerably larger than the proton radius, 
indicating that the dominant feature of DDIS is the probing with
the virtual photon of
non-perturbative exchanges similar to the pomeron 
of soft hadronic physics \cite{pomeron}. 

Fig.~\ref{regge}b shows the $Q^2$ dependence of the effective
pomeron intercept $\alpha_{I\!\!P} (0)$, which is extracted from 
the $x_{I\!\!P}$ dependence of the data \cite{zeus:lrglps}. 
No significant
dependence on $Q^2$ is observed, again compatible with
proton vertex factorization. 
These results are consistent with the H1 value of
%\begin{eqnarray}
$\alpha_{I\!\!P}(0) = 1.118 \pm 0.008 \ {\rm (exp.)} \
^{+0.029}_{-0.010} \ {\rm (model)}$ \cite{h1:lrg}.
%\end{eqnarray}
Both collaborations have also extracted a value for the slope of
the effective pomeron trajectory, the recently published ZEUS
value being 
%\begin{eqnarray}
$\alpha_{I\!\!P}^\prime = -0.01 \pm 0.06 \ {\rm (stat.)}
\ \pm 0.06 \ {\rm (syst.)} \ {\rm GeV^{-2}}$ \cite{zeus:lrglps}.
%\end{eqnarray}

The intercept 
of the effective pomeron trajectory
is consistent within errors with the `soft pomeron' results 
from fits to total cross sections and
soft diffractive data \cite{dl}.
%Again, it is to be expected that the lack of kinematic dependence
%of the effective intercept breaks down as $\beta \rightarrow 1$
%where higher twist contributions such as exclusive vector meson 
%production are present. 
Although larger effective intercepts 
have been measured in hard vector meson production, 
no deviations with either $Q^2$ or $\beta$ have yet been observed
in inclusive DDIS.   
The measured slope of the effective trajectory is smaller than the canonical
soft diffractive value of $0.25 \ {\rm GeV^{-2}}$ \cite{soft:alphaprime},
though it is compatible with results from the soft 
exclusive photoproduction of $\rho^0$ mesons at HERA \cite{blist}.
%KHOZE CLAIMS TO UNDERSTAND THIS - CITATION? 

\section{Diffractive Parton Density Functions}
\label{sec:dpdfs}

\begin{figure}[h]
\centerline{\includegraphics[width=0.4\textwidth]{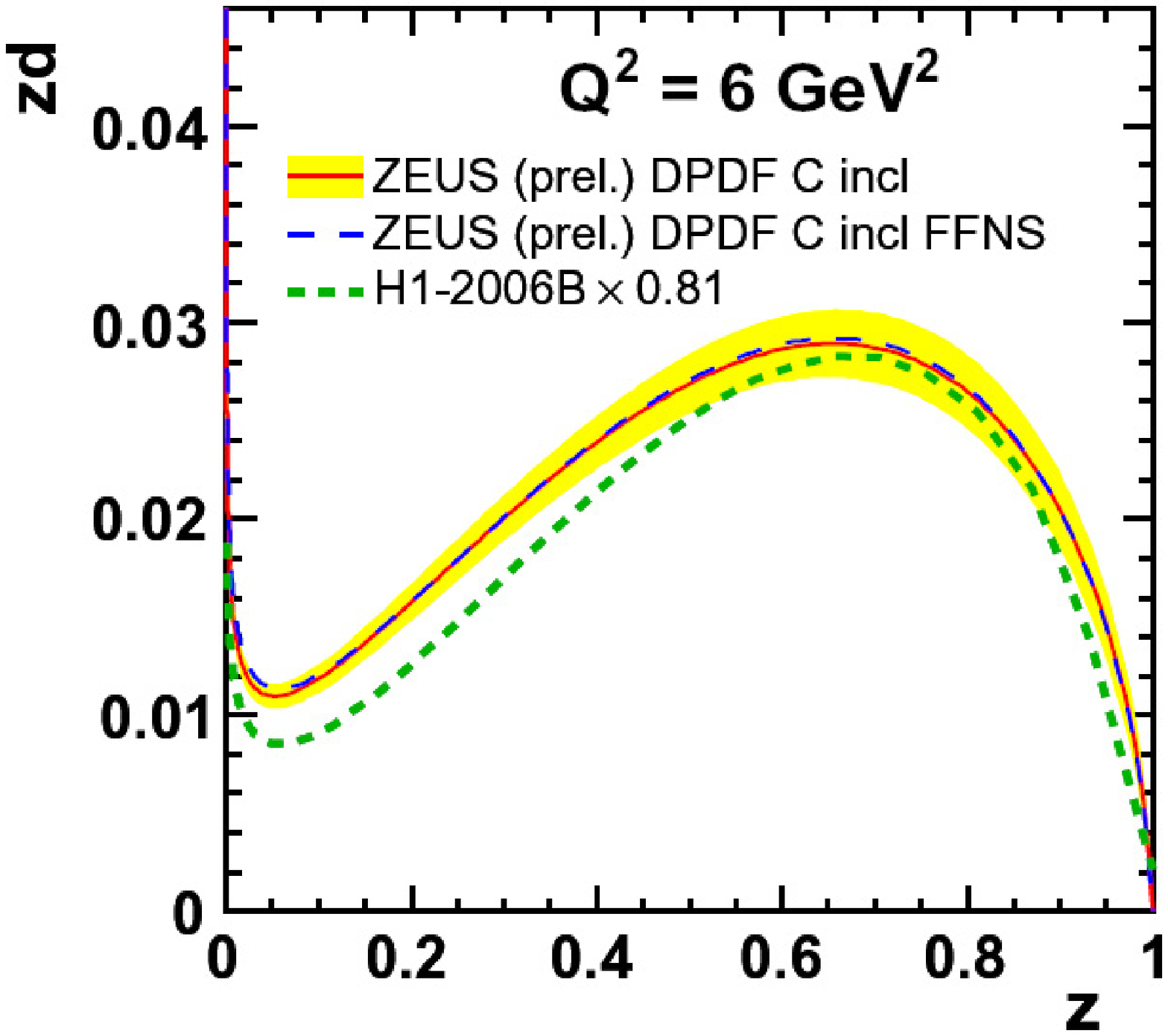}
            \hspace*{1cm}
            \includegraphics[width=0.3875\textwidth]{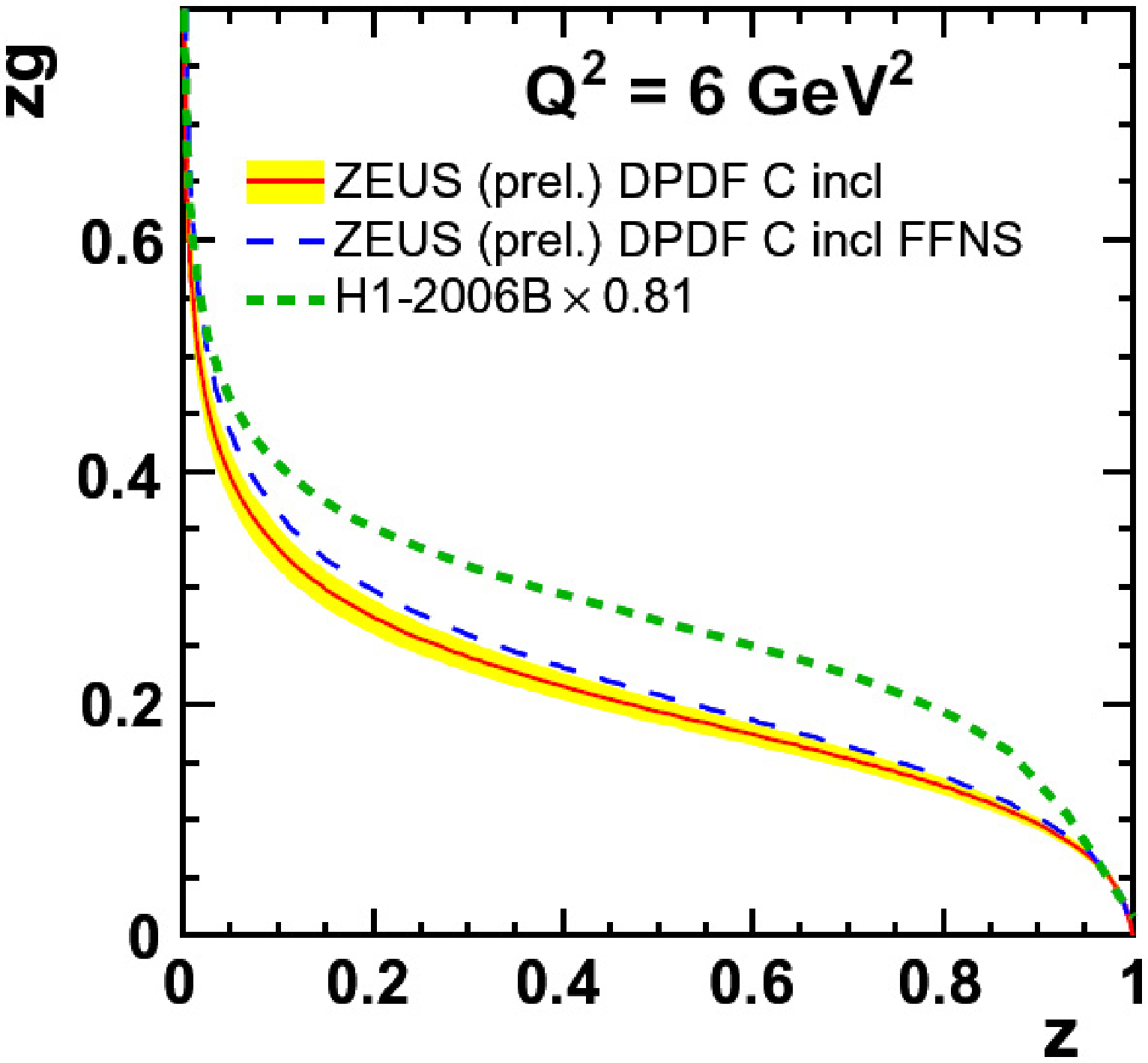}}
\caption{ZEUS down quark (one sixth of the total quark $+$ 
antiquark) and gluon densities
as a function of generalized momentum fraction $z$ at 
$Q^2 = 6 \ {\rm GeV^2}$ \cite{zeus:diffqcd}. 
Two heavy flavor schemes are shown, 
as well as H1 results \cite{h1:lrg} corrected
for proton dissociation with a factor of 0.81.}
\label{dpdfs}
\end{figure}

In the framework of the proof \cite{collins} of a
hard scattering collinear QCD factorisation theorem
for semi-inclusive DIS processes such as DDIS, the concept of `diffractive
parton distribution functions' (DPDFs)~\cite{facold}
may be introduced, representing conditional proton parton probability
distributions under the constraint of a leading final state proton with a
particular four-momentum.

The differential DDIS cross 
section may then be written in terms of convolutions of
partonic cross sections $\hat{\sigma}^{e i} (x, Q^2)$ with
DPDFs $f_i^D$ as
\begin{equation}
{\rm d} \sigma^{ep \rightarrow eXp} (x, Q^2, x_{_{I\!\!P}}, t) = \sum_i \
f_i^D(x, Q^2, x_{_{I\!\!P}}, t) \ \otimes \
{\rm d} \hat{\sigma}^{ei}(x,Q^2) \ .
\label{equ:diffpdf}
\end{equation}
The empirically motivated proton vertex factorization 
property (section~\ref{soft}) suggests a 
further factorization, whereby the DPDFs vary only
in normalization with the four-momentum of the final state proton
as described by $x_{_{I\!\!P}}$ and $t$:
\begin{equation}
f_i^D(x,Q^2,x_{_{I\!\!P}},t) = f_{I\!\!P/p}(x_{_{I\!\!P}},t) \cdot
f_i (\beta=x/x_{_{I\!\!P}},Q^2) \ .
\label{reggefac}
\end{equation}
Parameterizing $f_{I\!\!P/p}(x_{_{I\!\!P}},t)$
using Regge asymptotics, equation~\ref{reggefac}
amounts to a description of
DDIS in terms of the exchange of a factorisable
pomeron with universal parton densities \cite{ingelman:schlein}.
The 
$\beta$ and $Q^2$ dependences
of $\sigma_r^D$ may then be subjected to a perturbative QCD 
analysis based on the DGLAP equations in order to obtain DPDFs.
%similarly to the inclusive proton parton densities \cite{max}.
Whilst $F_2^D$ directly measures
the quark density, the gluon density is only indirectly
constrained, via the scaling violations 
$\partial F_2^D / \partial \ln Q^2$.

The high statistics ZEUS LRG and LPS data \cite{zeus:lrglps}
have recently been fitted to extract DPDFs \cite{zeus:diffqcd}.
The method and DPDF parameterization are similar to an earlier
H1 analysis \cite{h1:lrg}, the main step forward being in the heavy 
flavor treatment,
which now follows the general mass variable flavor number 
scheme \cite{gm:vfns}.
In Fig.~\ref{dpdfs},
the resulting DPDFs are compared with results from both ZEUS and H1 
using a fixed flavor number scheme. 
The agreement between the experiments
is reasonable when the uncertainty on the H1 DPDFs is also taken into
account and the conclusion that the dominant feature is a
gluon density with a relatively hard $z$ dependence is confirmed. 
The error bands shown 
in Fig.~\ref{dpdfs} represent experimental uncertainties only.
Whilst the quark densities are rather well known throughout the
phase space, the theoretical uncertainties
on the gluon density are large. Indeed, in the 
large $z$ region, where the 
dominant parton splitting is $q \rightarrow qg$, the
sensitivity of $\partial F_2^D / \partial \ln Q^2$ 
to the gluon density becomes poor
and different DPDF parameterizations 
lead to large variations  
\cite{h1:lrg,zeus:diffqcd}. Improved large $z$ constraints 
have been
obtained by including dijet data in the QCD 
fits \cite{matthias,zeus:diffqcd}.

In common with the inclusive proton
PDFs at low $x$, the DPDFs exhibit a
ratio of around 7:3 between gluons and quarks, consistent
with a common QCD radiation pattern far from the valence region. 

%-------------------------------------------------
\section{Nucleon tomography }
%-------------------------------------------------

Measurements of the  DIS (or DDIS) of leptons and nucleons, $e+p\to e+X$ (or $e+p\to e+X+Y$),
allow the extraction of Parton Distribution Functions (PDFs) (or diffractive PDFs) which describe
the longitudinal momentum carried by the quarks, anti-quarks and gluons that
make up the fast-moving nucleons. 
While PDFs provide crucial input to
perturbative QCD calculations of processes involving
hadrons, they do not provide a complete picture of the partonic structure of
nucleons \cite{Schoeffel:2009aa}. 
In particular, PDFs contain neither information on the
correlations between partons nor on their transverse motion.
Hard exclusive processes, in  which the
nucleon remains intact, have emerged in recent years as prime candidates to complement
this essentially one dimensional picture. 
The simplest exclusive process is the deeply virtual
Compton scattering (DVCS) or exclusive production of real photon, 
$e + p \rightarrow e + \gamma + p$.
This process is of particular interest as it has both a clear
experimental signature and is calculable in perturbative QCD. 
The DVCS reaction can be regarded as the elastic scattering of the
virtual photon off the proton via a colorless exchange, producing a 
real photon in the final state  \cite{dvcsh1,dvcszeus}. 
In the Bjorken scaling 
regime, 
QCD calculations assume that the exchange involves two partons, having
different longitudinal and transverse momenta, in a colorless
configuration. These unequal momenta or skewing are a consequence of the mass
difference between the incoming virtual photon and the outgoing real
photon. This skewness effect can
 be interpreted in the context of generalized
parton distributions (GPDs) \cite{qcd} and can bring new insights
on the quarks/gluons imaging of the nucleon.

\begin{figure}[!]
\begin{center}
\includegraphics[width=6cm,height=4.cm]{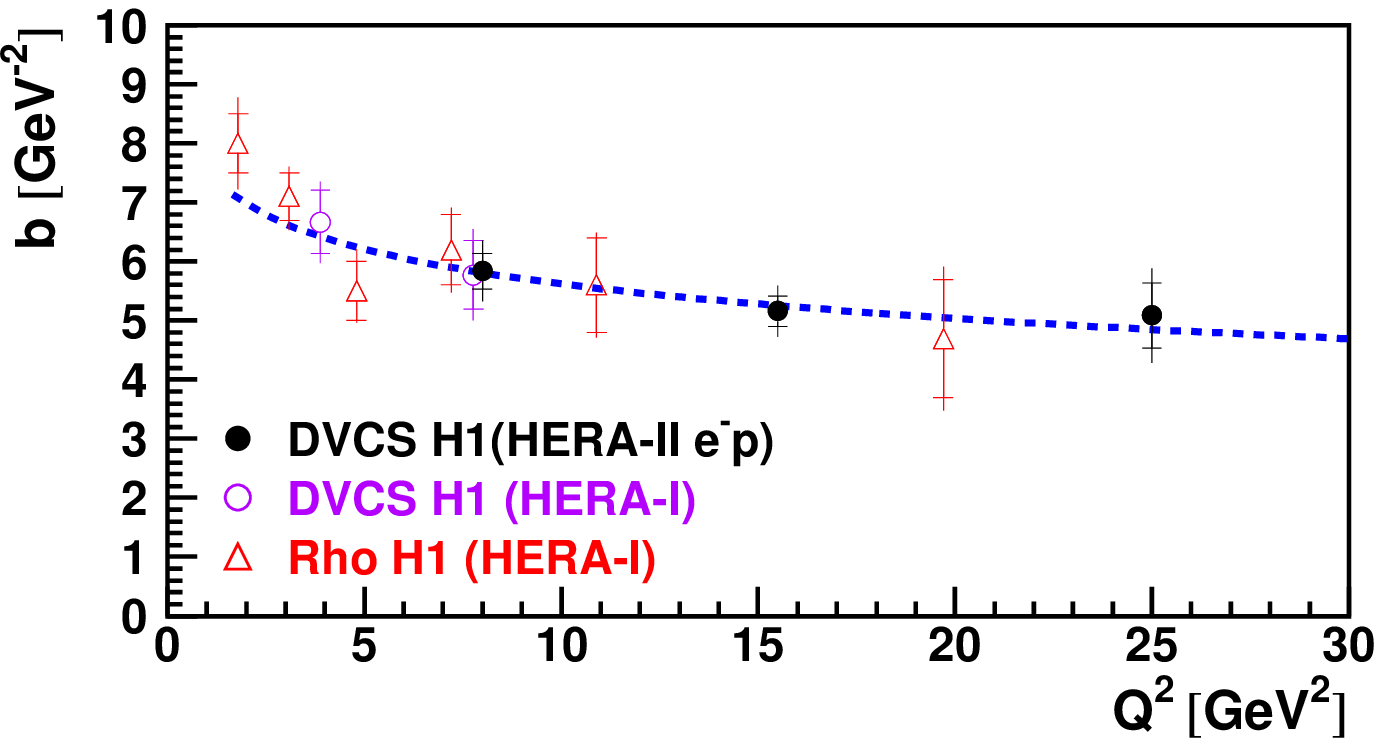}
\includegraphics[width=6cm,height=4.cm]{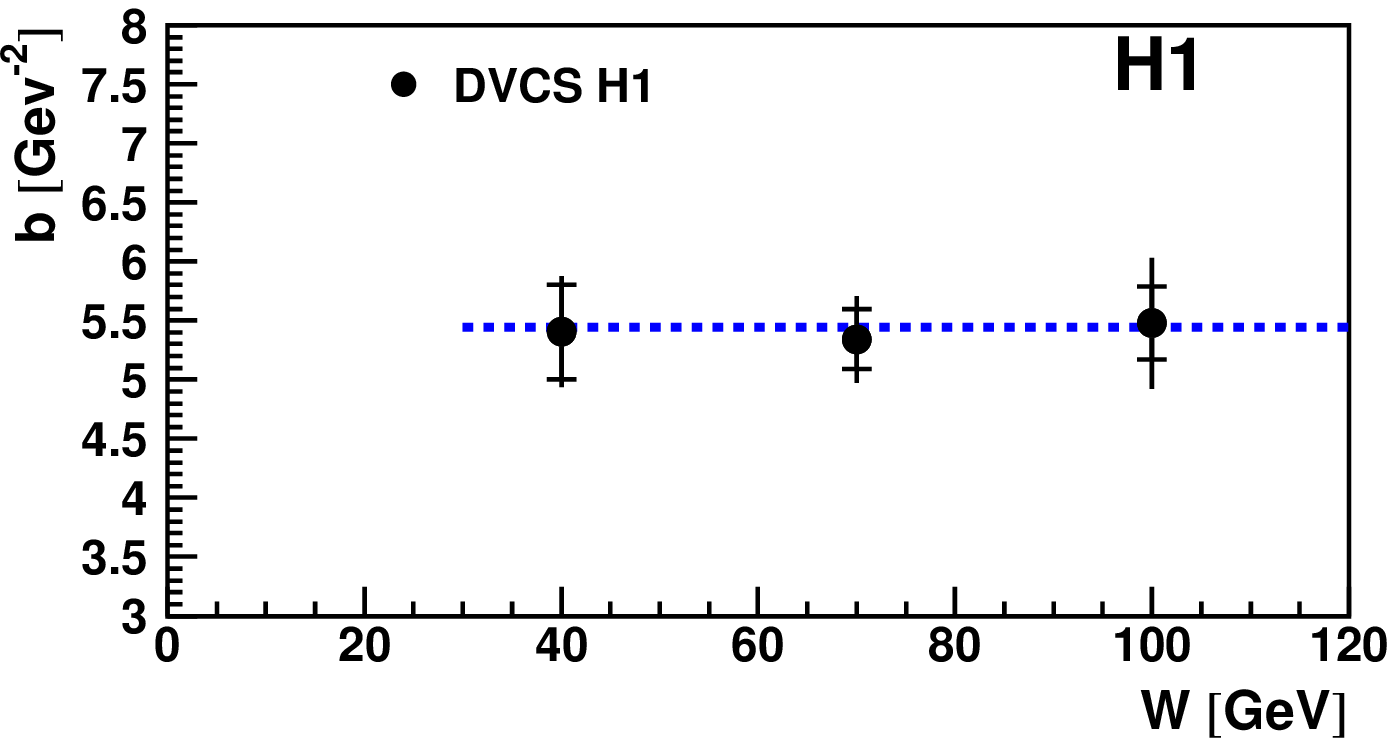}
\caption{ The logarithmic slope of the $t$ dependence
  for DVCS and $\rho$ exclusive production :
  $d\sigma/dt \propto
\exp(-b|t|)$  where $t=(p-p')^2$.
}
\label{bslopes}
\end{center}
%%\vspace{-0.7cm}
\end{figure}

One of the key measurement in exclusive processes is the slope
defined  by  the exponential fit to the differential cross section:  
$
d\sigma/dt \propto
\exp(-b|t|)
$
at small $t$, where $t=(p-p')^2$ is the square of the momentum transfer at the
proton vertex (see Fig. \ref{bslopes}). 
A Fourier transform from momentum
to impact parameter space readily shows that the $t$-slope $b$ is related to the
typical transverse distance between the colliding objects \cite{buk,diehl}.
At high scale, the $q\bar{q}$ dipole is almost
point-like, and the $t$ dependence of the cross section is given by the transverse extension 
of the gluons (or sea quarks) in the  proton for a given $x_{Bj}$ range.
More precisely, from the  generalized gluon distribution $F_g$ defined in section 3, we can compute
a gluon density which also depends on a spatial degree of freedom, the transverse size (or impact parameter), labeled $R_\perp$,
in the proton. Both functions are related by a Fourier transform 
$$
g (x, R_\perp; Q^2) 
\;\; \equiv \;\; \int \frac{d^2 \Delta_\perp}{(2 \pi)^2}
\; \exp[{i (\Delta_\perp R_\perp)}]
\; F_g (x, t = -{\Delta}_\perp^2; Q^2).
$$

Thus, the transverse extension $\langle r_T^2 \rangle$
 of gluons (or sea quarks) in the proton can be written as
$$
\langle r_T^2 \rangle
\;\; \equiv \;\; \frac{\int d^2 R_\perp \; g(x, R_\perp) \; R_\perp^2}
{\int d^2 R_\perp \; g(x, R_\perp)} 
\;\; = \;\; 4 \; \frac{\partial}{\partial t}
\left[ \frac{F_g (x, t)}{F_g (x, 0)} \right]_{t = 0} = 2 b
$$
where $b$ is the exponential $t$-slope.
Measurements of  $b$
have been performed  for  different channels, as DVCS or $\rho$ production (see Fig. \ref{bslopes}-left-),
which corresponds to $\sqrt{r_T^2} = 0.65 \pm 0.02$~fm at large scale $Q^2$ for $x_{Bj} \simeq 10^{-3}$.
This value is smaller that the size of a single proton, and, in contrast to hadron-hadron scattering, 
it does not expand as energy $W$ increases
(see Fig. \ref{bslopes}-right-).
This result is consistent with perturbative QCD calculations in terms of a radiation cloud of gluons and quarks
emitted around the incoming virtual photon.

\subsection{  Link with LHC issues }

%%%%%%%%%%%%%%%%%%%%%%%%%%%%%%%%%%%%%%%%%%%%%
\begin{figure}[htbp]
\begin{center}
  \includegraphics[width=0.34\textwidth]{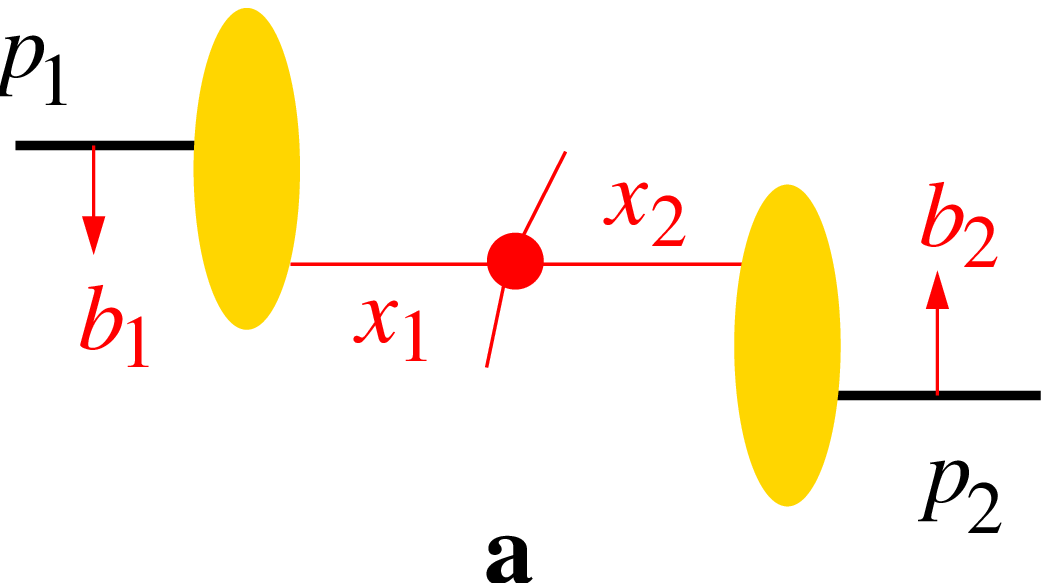}
  \hspace{0.1\textwidth}
  \includegraphics[width=0.34\textwidth]{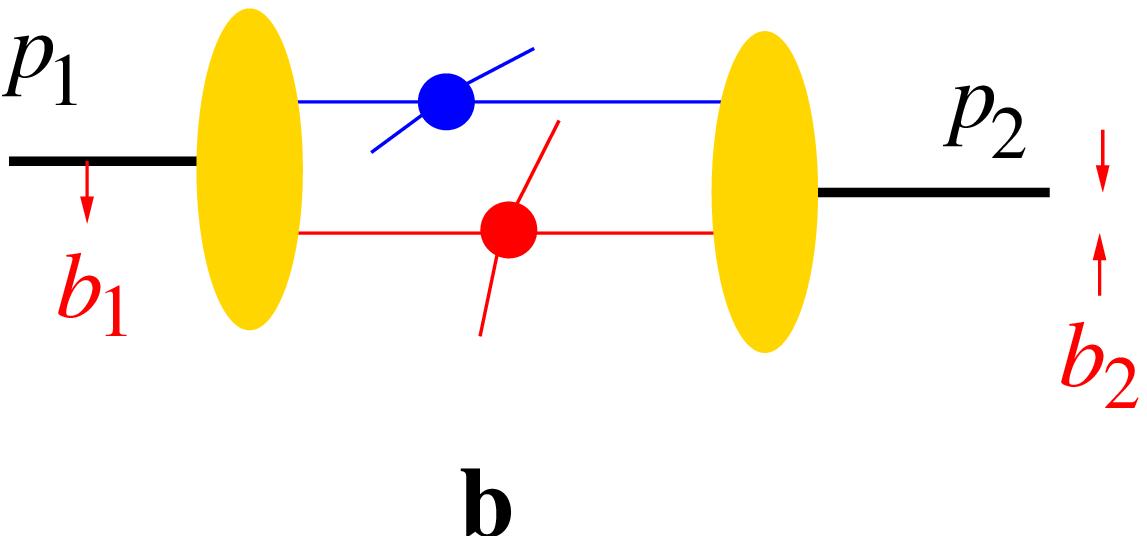}
\caption{\label{fig:multi}  a: Graph with a single hard
  interaction in a hadron-hadron collision.  The impact parameters
  $b_1$ and $b_2$ are integrated over independently. b:
  Graph with a primary and a secondary interaction. }
  \end{center}
\end{figure}

The  correlation between the transverse distribution of partons
and their momentum fraction is not only interesting from the
perspective of hadron structure, but also has practical consequences
for high-energy hadron-hadron collisions.
Consider the production of a high-mass system (a dijet or a
heavy particle).  For the inclusive production cross section, the
distribution of the colliding partons in impact parameter is not
important: only the parton distributions integrated over impact
parameters are relevant according to standard hard-scattering
factorization (see Fig.~\ref{fig:multi}(a)).  There can however be
additional interactions in the same collision, especially at the high
energies for the Tevatron or the LHC, as shown in
Fig.~\ref{fig:multi}(b).  Their effects cancel in sufficiently
inclusive observables, but it does affect the event
characteristics and can hence be quite relevant in practice.  In this
case, the impact parameter distribution of partons must be
considered. 

The
production of a heavy system requires large momentum fractions for the
colliding partons.  A narrow impact parameter distribution for these
partons forces the collision to be more central, which in turn
increases the probability for multiple parton collisions in the event
(multiple interactions).

%-------------------------------------------------
\section{Conclusions }
%-------------------------------------------------

We have presented and discussed the most recent results on
 diffraction from the HERA experiments, H1 and ZEUS.
Inclusive diffraction have been shown to be closely related to the high gluon density in the proton.
With exclusive processes studies, we have illustrated the importance of  $t$-slope measurements 
in order to get a better understanding of how quarks and gluons are assembled in the nucleon.

%=========================================================================

\end{document}